\documentclass[runningheads]{cl2emult}
\usepackage{graphicx}
\usepackage{latexsym}

\begin{document}
\title*{Experimental observations of non-equilibrium distributions and
transitions in a 2D granular gas
}
\titlerunning{Non-equilibrium distributions and
transitions in a 2D granular gas}
%
\author{Jeffrey S. Urbach 
\and Jeffrey S. Olafsen}

\institute{Department of Physics, Georgetown University, Washington, DC 20057}

\maketitle              

\begin{abstract}
A large number
($\sim$10,000) of uniform stainless steel balls comprising less than one
layer coverage on a vertically shaken plate provides a
rich  system for the study of excited granular media. Viewed
from above, the horizontal motion in the layer shows interesting
collective behavior as a result of inelastic particle-particle
collisions. Clusters appear as localized fluctuations from purely random
density distributions, as demonstrated by increased particle
correlations. The clusters grow as the medium is "cooled" by reducing
the rate of energy input. Further reduction of the energy input leads to
the nucleation of a collapse: a close-packed crystal of particles at
rest. High speed
photography allows for measurement of particle velocities between
collisions. The velocity distributions deviate strongly from a Maxwell
distribution at low accelerations, and show approximately exponential
tails, possibly due to an observed cross-correlation between density and
velocity fluctuations. When the layer is confined with a lid, the
velocity
distributions at higher accelerations are non-Maxwellian and
independent of the granular temperature.
\end{abstract}

\section{Introduction}
The development of a kinetic theory of granular gases, collections of
large numbers of inelastically colliding particles, has proven to be a
very challenging undertaking.  While the equilibrium properties of
elastically colliding gases are relatively well understood, the
introduction of physically relevant levels of dissipation change the
dynamics dramatically.  Considerable progress has been made in
understanding the behavior of freely cooling granular gases, where the
energy lost to collisions is not replaced.  In order to model a
variety of industrial processes where energy is added to the grains to
enhance mixing and transport, a kinetic theory of forced granular
gases is required. By analogy with the theory of equilibrium fluids,
the goal is to solve a Boltzmann-like equation for the relevant
`microscopic' statistical distribution functions describing the
behavior of individual grains. The rate at which energy is put into
the fluctuating velocities is balanced by the rate at which it is lost
in collisions.  `Macroscopic' transport coefficients, relating average
fluxes (or currents) to gradients in local (coarse grained) variables,
can then be calculated from the appropriate correlation functions.

Some progress has been made in building a continuum description of
granular fluids from a microscopic kinetic theory, but
has typically involved a number of approximations that have yet to be tested. 
A hydrodynamic
description is only possible if the spatial and temporal fluctuations
in the flow are sufficiently localized to permit coarse graining. This
may be the case if the granular system is large
enough and the energy in the flow is high enough, but making this
statement quantitative is very much an ongoing effort.
Precise
experimental measurements of the steady state statistical
properties in a variety of granular fluids are essential to
developing and testing generally applicable descriptions.  We
have been investigating the fluidized state and 
the fluidization transition in a simple representation of a granular
material: a single layer of identical spherical particles.  Using
identical particles simplifies analysis of the system and allows
for comparison with experimental and theoretical results derived
for atomic and molecular dynamics.  The use of a single granular
layer, in combination with high speed imaging technology, allows for a
thorough description of the granular system, including particle
velocity distributions, correlation functions, and transport
properties.  We have
found a complex phase diagram that bears many similarities to
equilibrium two-dimensional systems, but we have also  directly measured
particle velocity distributions that are non-Maxwellian, 
and density correlation
functions that show non-equilibrium effects \cite{olafsenprl}.  
We have also measured the cross-correlation between density and
temperature fluctuations, and the effects of constraining the granular
layer by placing a lid above it \cite{olafsenpre}.

This paper is organized as follows: Section \ref{setup} describes the
experimental setup and analysis techniques, and Section \ref{phases}
describes the experimentally observed phase diagram.  Our results on
the statistical characterization of the granular gas are reported in
\ref{gas}, followed by a comparison with related work and a discussion of future directions 
in Section \ref{discussion}.


\section{Experimental setup and methods}
\label{setup}
The experimental apparatus consists of a 20 cm diameter smooth, rigid
aluminum plate that is mounted horizontally on an electromagnetic
shaker that oscillates the plate vertically.  The plate is carefully
leveled, and the amplitude of acceleration is uniform across the plate
to better that 0.5 \%.  The acceleration of the plate is monitored
with a fast-response accelerometer mounted on the bottom surface of
the plate. Two types of particles were used for the experiments
described below: smaller spheres of 302 stainless steel with an
average diameter of 1.191 $\pm$ 0.0024 mm, and larger spheres
of 316 stainless with an average diameter of 1.588 $\pm$ 0.0032
mm.  The coefficient of restitution, both particle-particle and
particle-plate, is about 0.9.  The particles are contained by an
aluminum rim that occupies the outer 1.9 cm of the plate. 
The particles are illuminated by low
angle diffuse light.  This illumination produces a small bright spot
at the top of each particle when viewed through a video camera mounted
directly above the plate.

Two digital video cameras are used for data acquisition, a
high-resolution camera for studying spatial correlations (Pulnix-1040,
1024 x 1024 pixels, Pulnix America, Inc., Sunnyvale, CA), and a
high-speed camera for measuring velocity distributions (Dalsa CAD1,
128 x 128 pixels, 838 frames/second, Dalsa Inc., Waterloo, ON;
Canada).  A collection of images and movies is available for viewing
at www.physics.georgetown.edu/$\sim$granular.  The acquired digital
images are analyzed to determine particle locations by calculating
intensity weighted centers of bright spots identified in the images.

The frame rate of the high speed camera is much faster than
the inter-particle collision rate, so it is possible to measure
particle velocities by measuring the displacement of the particle from
one image to the next.  The trajectories determined from the images
are nearly straight lines between collisions.  Although the frame rate
is fast compared the collision time, the fact that it is finite
introduces unavoidable systematic errors into the experimentally
determined velocity distributions.  Some particles will undergo a
collision between frames, and the resulting displacement will not
represent the true velocity of the particle either before or after the
collision.  Since fast particles are more likely to undergo a
collision, the high energy tails of the distribution will be
decreased.  The probability that a particle of velocity $v$ will not
undergo a collision during a time interval $\Delta t$ is proportional
to exp$(-v \Delta t / l_o)$, where $l_o$ is the mean free path.  The
effect of this on the velocity distribution function reported here is
quite small.  In addition to taking velocities out of the tails of the
distribution, collisions will incorrectly add the velocities of those
particles elsewhere in the distribution.  This effect can be minimized
by filtering out the portion of particle trajectories where collisions
occur.  This is done by analyzing particle position in 3 images at a
time.  When the change in apparent velocity from the first two images
differs from the change in the second two by more than a specified
cutoff value, the points are neglected.  Because a relatively small
fraction of the particles undergo a collision between any two frames,
the results are not sensitive to the precise value chosen.  At the
highest accelerations considered in this work, varying the cutoff
value over more than an order of magnitude changes the measured
granular temperature ({\it rms} velocity) by less than 10\%, and the
flatness of the distribution (Eq. \ref{eq:flatness}) 
by less than 0.2.
Similar issues are considered in a somewhat different manner in
reference \cite{LosertCDKG99}.


\section{Phase diagram}
\label{phases}
\begin{figure}[h]
\centering
\includegraphics[width=1.0\textwidth]{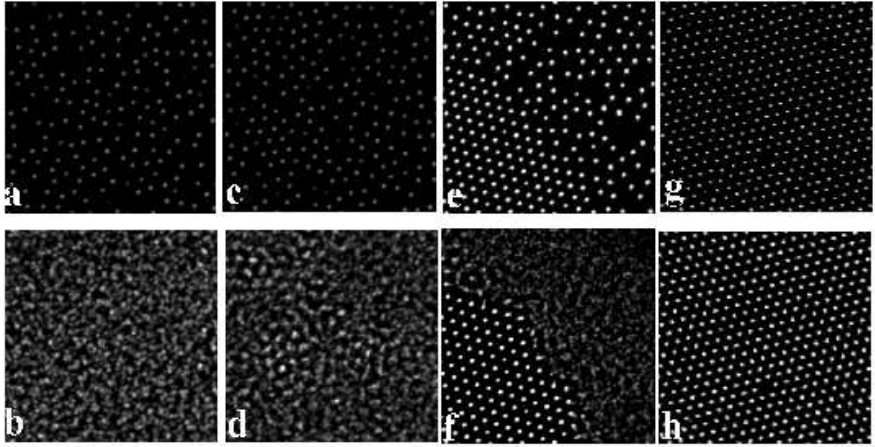}
\caption{Instantaneous (top row) and time-averaged (bottom row) 
photographs detailing the different phases of the granular 
monolayer.  (a), (b), uniform particle distributions typical of the gas phase
($\rho = 0.463, \Gamma = 1.01, \nu = 70$ Hz).  
(c) The clustered phase ($\rho = 0.463, \Gamma = 0.8, \nu = 70$ Hz).  
The higher intensity points in a time-averaged image, (d),
denote slower, densely packed particles.  
(e) A portion of a collapse for $\rho = 0.463$, $\Gamma = 0.76$, and
$\nu = 70$ Hz.
(f) The time-averaged image shows that the particles in the collapse are 
stationary while the surrounding gas particles continue to move.
(g),(h) In a more dense system, $\rho = 0.839$, there is an ordered phase
($\Gamma = 1.0, \nu = 90$ Hz) where all of the particles remain in motion.}
\label{fig:phases}
\end{figure}

When the plate oscillation amplitude is not too large, the spheres
never hop over one another; thus the system essentially
two-dimensional.  Nonetheless, there is sufficient energy in the
horizontal velocity component to generate fascinating dynamic
phenomena.  At moderately large sinusoidal vibration amplitudes, a
fully fluidized state is observed. The spheres are constantly in
motion and there is no large-scale spatial ordering.  Figure
\ref{fig:phases}(a) shows an instantaneous image of part of a cell
containing 8000 particles  in this
regime. (The number of particles in a single hexagonal close-packed
layer is $N_{max}=17275$ for the smaller particles, giving a reduced
density $\rho=N/N_{max}=0.463$.) The peak plate acceleration relative to the acceleration due
to gravity, $\Gamma = A \omega ^2 / g$, is just over one.  This phase
is characterized by an apparently random distribution of particle
positions and velocities.  A sense of the dynamics can be gained from
an average of 15 frames taken over a period of 1 s
(Fig. \ref{fig:phases}(b)), which shows the lack of any stable
structure. As the amplitude of the acceleration is slowly decreased,
the average kinetic energy of the particles decreases, and localized
transient clusters of low velocity particles appear.  An instantaneous
image in this regime (Fig. \ref{fig:phases}(c)) does not look very
different from the one taken at higher acceleration, but in the
time-averaged image (Fig. \ref{fig:phases}(d)) bright peaks are
clearly evident, corresponding to low-velocity particles that have
remained relatively close to each other over the time interval.  In
this regime, the clusters typically survive for 1-20 seconds.  In the
low density regions outside of the clusters, there are particles with
anomalously high velocities, and these appear to be responsible for
the breakup of the clusters.  There are no attractive interactions
between these particles; the cluster formation is a uniquely
non-equilibrium phenomenon, resulting from the dissipation during
inter-particle collisions, and the particle-plate dynamics.

When the amplitude of the vibration is decreased somewhat below that
of Fig. \ref{fig:phases} (c,d), the typical cluster size increases to
12-15 particles. Within a few minutes at this acceleration, one of
these large clusters will become a nucleation point for a `solid'
phase, similar to what is referred to as 'inelastic collapse'
\cite{McNamaraY96}.
The
particles in the collapse are in contact with all of their neighbors,
and form a perfect hexagonal lattice (Fig. \ref{fig:phases}(d)).  The
collapse is surrounded by a gas of the remaining particles.  The sharp
interface between the coexisting phases can be seen in the
time-averaged image (lower panel of Fig. \ref{fig:phases}(e)) .  The
two-phase co-existence persists essentially unchanged for as long as
the driving is maintained.  At higher densities, instead of a
transition directly from the clustering behavior to collapse, there is
an intermediate phase with apparent long range order.  Figure
\ref{fig:phases}(g) shows a monolayer in this ordered state, where the
spheres are arranged in a hexagonal lattice but are not at rest or in
contact with one another.  The disorder in the image is a consequence
of the fluctuations induced by inter-particle collisions.  When the
particle positions in the ordered phase are averaged over a short time
(Fig. \ref{fig:phases}(h)), the resulting image displays a nearly
perfect lattice, with one unoccupied site. 
Measurements of the correlation functions for
positional and orientational order parameters in this phase suggest
that the transition to this ordered phase is quantitatively similar to
the liquid-solid transitions observed in a variety of equilibrium
systems.

\begin{figure}[h]
\centering
\includegraphics[width=.8\textwidth]{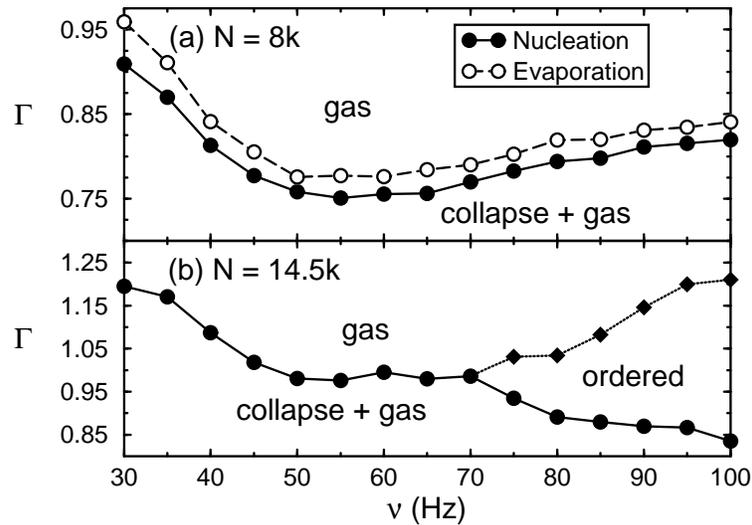}
\caption{The phase diagrams for (a) N = 8000 particles ($\rho = 0.463$)
and (b) N = 14,500 ($\rho = 0.839$)
particles.  The filled circles denote the acceleration where the collapse
nucleates.  The open circles in (a) indicate the point where the collapse
disappears upon increasing the acceleration.  The diamonds in 
(b) show the transition to the ordered state as the acceleration is reduced.}
\label{phasediagram} 
\end{figure}

Figure \ref{phasediagram} shows phase diagrams of the system with two
different densities.  The data for the nucleation points were taken by
decreasing the plate acceleration in steps of about 0.003$g$, and
waiting 5 minutes at each step to see if the collapse nucleates. The
precise location of the nucleation line depends on the waiting time,
but only very weakly. Once the collapse forms, increasing the
acceleration causes it to 'evaporate' as particles return to the gas
phase.  Th open circles in Figure \ref{phasediagram}(a) indicate the
acceleration required to fully evaporate the collapse. 
The evaporation line is omitted from the lower panel
for clarity.   

\begin{figure}[h]
\centering
\includegraphics[width=.6\textwidth]{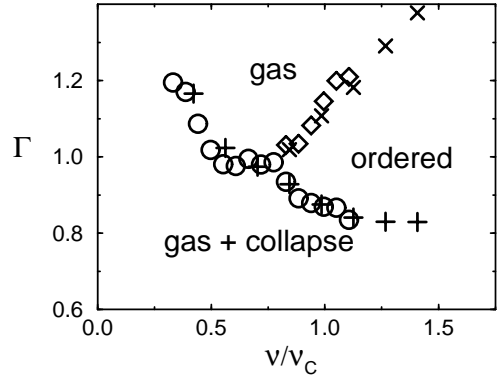}
\caption{ The phase diagrams for 1.2 mm diameter particles 
($\bigcirc$, $\diamond$) and 1.6 mm ($+$,$\times$) at $\rho=0.839$.  The frequency is scaled by $\nu _c = (g/d)^{1/2}$. }
\label{overlay} 
\end{figure}

It is not immediately clear what causes the
frequency dependence in the phase diagram.  For an ideal spherical
particle on an oscillating plate with a velocity-independent
coefficient of restitution, the dynamical behavior depends only on
$\Gamma$, and the frequency sets the timescale for the motion, and the
length scale through $g / \omega ^2$, which is proportional to the
distance a ball falls during one oscillation period.  Thus as the
frequency is reduced for fixed $\Gamma$, balls will bounce higher.
Because the balls in this system interact with their neighbors, it is
possible that the frequency dependence enters through the ratio of
this length scale to the ball diameter.  In fact, the rapid increase in
the acceleration where collapse forms for frequencies below about 50
Hz (see Fig. \ref{phasediagram}) occurs when the particles begin to
bounce high enough to hop over one another, resulting in a gradual
transition from primarily 2D to 3D dynamics.  This suggests that the
frequency dependence comes from a characteristic frequency $\nu _c =
(g/d)^{1/2}$, where $d$ is the sphere diameter.  Figure \ref{overlay}
shows the phase diagram measured for two different sets of particles
determined at the same reduced density $\rho$,
one with a diameter of 1.2 mm, and one with a diameter of 1.6 mm.  The
upturn at low frequencies and the appearance of the ordered phase
occur at lower frequencies for the larger particles, but the scaled
phase diagrams lie right on top of one another.

\section{Statistical characterization of the granular gas}
\label{gas}
In order to incorporate the non-equilibrium fluctuations observed in
Fig. \ref{fig:phases}(c,d) into a kinetic theory of the granular fluids, 
quantitative measures are required.  In the monolayer
system, particle positions can be directly determined from
images acquired with a digital video camera.  These can be used to evaluate
statistical measures, such as the pair correlation function and the velocity 
distribution function, that are essential components of kinetic theory.
\subsection{Pair Correlation Functions}
\label{position}
Correlations in particle positions are most directly measured by the pair
correlation function, $G(r)$:

\begin{equation}
G(r) = \frac{<\rho(0) \rho(r)>}{<\rho>^2},
\label{eq:two}
\end{equation}

\noindent
where $\rho$ is the particle density. The correlations of a
two-dimensional gas of elastic hard disks in equilibrium 
are due only to geometric
factors of excluded volume and are independent of temperature
\cite{LandauL}.

\begin{figure}[h]
\centering
\includegraphics[width=.6\textwidth]{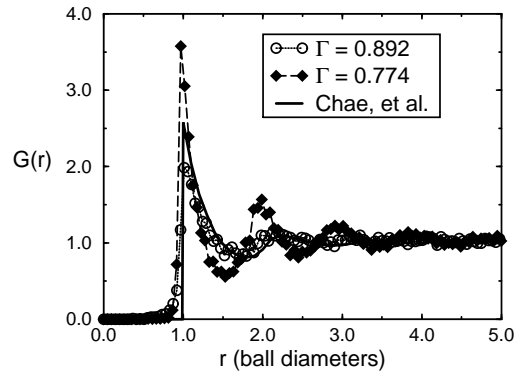}
\caption{\small{ The measured pair correlation function of a
two-dimensional granular gas ($\rho = 0.463, \nu = 70$ Hz).  The results
are compared to the result from an equilibrium hard
disk Monte Carlo calculation with no adjustable parameters. The
legend gives the value of $\Gamma$.)
 }}
\label{fig3}
\end{figure}

The solid line in Fig. \ref{fig3} shows $G(r)$ from a Monte Carlo
calculation of a two-dimensional gas of elastic hard disks in
equilibrium for a density of 0.463 \cite{ChaeRR69}.  The
experimentally measured correlation function in the gas-like phase,
shown by the open circles, is almost identical to the equilibrium
result. There are no free parameters in Fig. \ref{fig3}.  The
agreement between the experimental correlation function and the
equilibrium result suggests that the structure in the correlation
function of the gas-like phase is dominated by excluded volume
effects.  As the granular medium is cooled, the correlations grow
significantly.  This is evident from the data for a lower vibration
amplitude ($\Gamma$ = 0.774, just above the acceleration where the
collapse forms), shown as filled diamonds in Fig. \ref{fig3}.  The
increased correlations indicate that there are non-uniform density
distributions in the medium: high density regions of relatively
closely packed particles, which must coexist with relatively low
density regions.  In the kinetic theory of equilibrium fluids, 
the pair correlation
function plays a central role, because thermodynamic quantities can be
written as integrals with $G(r)$. The equation of state for an
equilibrium hard disk system, as well as the Enskog modification to
the Boltzmann equation (the correction to account for excluded volume
effects), depends on $G(r)$ at contact ($r=$1
diameter) \cite{hansen}. Extensions of kinetic theory to inelastic
gases \cite{jenkins} assume that $G(r)$ at contact is given by the
Carnahan-Starling relationship \cite{cs}, which works well for elastic
gases.  Thus the dramatic increase in correlations will likely have
significant quantitative implications for transport coefficients in
the granular gas.

\begin{figure}[h]
\centering
\includegraphics[width=.6\textwidth]{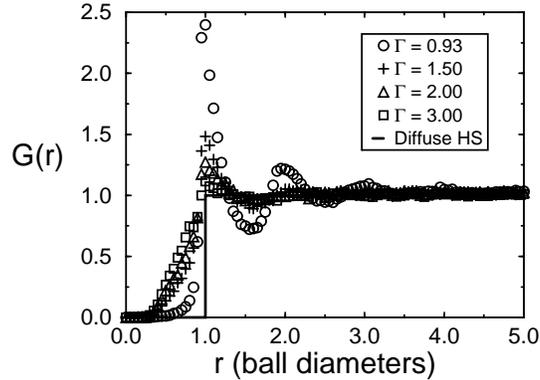}
\caption{(a) Pair correlation functions for larger accelerations in the unconstrained system. ($\bigcirc$) $\Gamma = 0.93$,
($+$) $\Gamma = 1.5$, ($\triangle$) $\Gamma = 2.0$, ($\Box$) $\Gamma =
3.0$.  }
\label{highpair}
\end{figure}

At higher plate accelerations, the pair correlation function loses all
of its structure.  Fig. \ref{highpair} shows $G(r)$ at $\Gamma =
0.93$, $\Gamma = 1.50$, and $\Gamma = 3.0$.  Increasing the steady
state kinetic energy of the granular gas by increasing the amplitude
of the acceleration at constant frequency causes the gas to change
from primarily two-dimensional, where the particles never hop over one
another, to essentially three-dimensional.  This
transition can be observed in the pair correlation function, $G(r)$,
by the increase in its value for $r<1$. (The correlation function
includes only horizontal particle separations.)  This
transition can affect the dynamics in several ways: the effective
density is decreased, so that excluded volume effects are less
important; the inter-particle collisions can occur at angles closer to
vertical, affecting transfer of energy and momentum from the vertical
direction to the horizontal; and the change in the dimensionality
itself can have important consequences.

In order to separate these effects
from the direct consequences of increasing the kinetic energy of the gas,
a Plexiglas lid was added to the system at a height of 2.54 mm, or 1.6
ball diameters for the larger particles.  For this plate-to-lid separation,
the larger particles cannot pass over top of one another, although
enough room remains for collisions between particles at sufficiently
different
heights to transfer momentum from the vertical to the horizontal direction.

\begin{figure}[h]
\centering
\includegraphics[width=.6\textwidth]{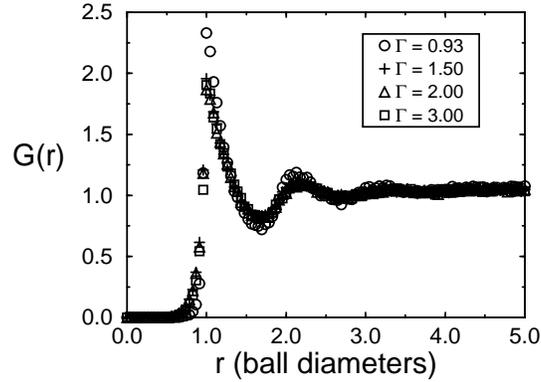}
\caption{ Pair correlation functions for the velocity distributions where a lid
constrains the system to remain two dimensional.  The particle
correlations remain as the shaking amplitude is increased.
($\bigcirc$) $\Gamma = 0.93$,
($+$) $\Gamma = 1.5$, ($\triangle$) $\Gamma = 2.0$, ($\Box$) $\Gamma = 3.0$.}
\label{lidpair}
\end{figure}

Figure \ref{lidpair} shows the pair correlation functions measured
with the lid on, and demonstrates that the particle-particle
correlations persist and become independent of $\Gamma$ when the
system is constrained in the vertical direction.  The correlation
function decreases slightly from $\Gamma = 0.93$ to $\Gamma = 1.50$,
and then remains essentially constant up to $\Gamma = 3$.  The small
value of $G(r)$ for distances less than one ball diameter indicates
that the system remains 2D as $\Gamma$ is increased.  The structure
observed in the correlation function is essentially the same as that
of an equilibrium elastic hard sphere gas at the same density,
indicating that the correlations that exist are due to excluded volume
effects.  From $\Gamma = 1.50$ to $\Gamma = 3.0$, the granular
temperature changes by more than a factor of 2 (see Fig. \ref{tvsg}), yet
there is no detectable change in the pair correlation function.  Thus,
unlike the low acceleration behavior shown in Fig \ref{fig3}, the
particle correlations in the constrained system at higher
accelerations behave like those of a system of elastic
hard disks.

\subsection{Velocity Distribution Functions}
\label{velocity}

A crucial ingredient of a statistical approach to describing the
dynamics in a granular system is the velocity distribution, which may
show non-equilibrium effects as can the correlation function.  As
described in section \ref{setup}, the horizontal components of the
particle velocities between collisions can be determined with the use
of a high speed camera.  Extensive measurement of the velocity distributions
in the plane of the granular gas demonstrate non-Gaussian behavior.

\begin{figure}[h]
\centering
\includegraphics[width=.7\textwidth]{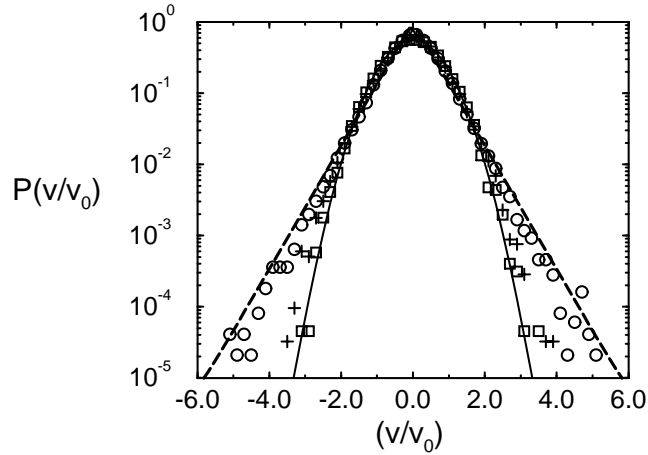}
\caption{ Velocity distributions for the system without a lid.  
As the acceleration is increased, the distributions go from having nearly 
exponential tails to Gaussian tails.
($\bigcirc$) $\Gamma = 0.93$,
($+$) $\Gamma = 1.5$, ($\Box$) $\Gamma = 3.0$.}
\label{veldist}
\end{figure}

The measured horizontal velocity distributions at $\Gamma = 0.93$,
$\Gamma = 1.50$, and $\Gamma = 3.0$ are shown in
Fig. \ref{veldist}. The distributions at low $\Gamma$ are strongly
non-Gaussian, showing approximately exponential tails
\cite{olafsenprl}.  As the acceleration is increased, the distribution
crosses over smoothly to a Gaussian.  This behavior is superficially
similar to that of freely cooling granular media, where an initial
Gaussian velocity distribution becomes non-Gaussian as the system
cools, but these results are obtained in a steady state.

The non-Gaussian velocity distributions observed at 
low accelerations are accompanied by clustering, as demonstrated by a 
dramatic increase in the structure of the pair correlation function.  
Conversely, the crossover
to Gaussian velocity distributions is accompanied by the disappearance 
of spatial correlations, consistent with the suggestions that the 
non-Gaussian velocity distributions arise from a coupling between density
and temperature fluctuations \cite{puglisi}.

\begin{figure}[h]
\centering
\includegraphics[width=.7\textwidth]{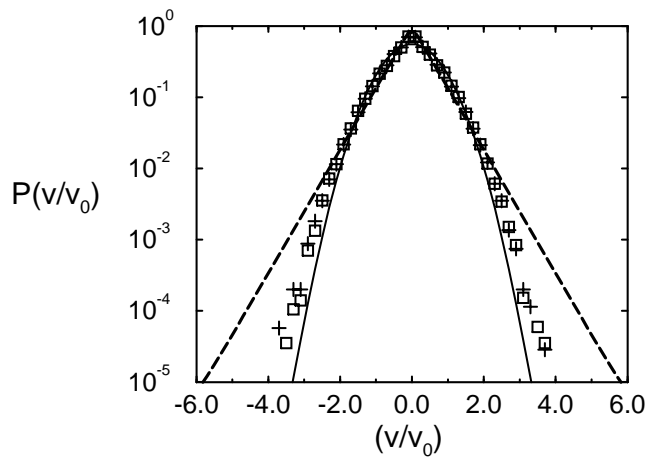}
\caption{ Velocity distributions for the system with a lid.  
The scaled distributions are essentially independent of acceleration.
($+$) $\Gamma = 2.0$,
($\Box$) $\Gamma = 3.0$.}
\label{lidveldist}
\end{figure}

The velocity distributions measured with the lid on are shown in
Figure \ref{lidveldist}.  The crossover to Gaussian distributions
observed when the acceleration is increased in the unconstrained
system is not observed.  Instead, like the pair correlation function,
the statistical characteristics of the granular gas become independent
of acceleration, and therefore independent of the granular temperature.
The tails of the distribution in the constrained system are consistent
with $P(v)
\propto$ exp$(-|v|^{3/2})$, as observed in \cite{LosertCDKG99} and predicted by
\cite{vannoije}. 
The relationship between our results and that work is discussed in
section \ref{ggdiscussion}.

\begin{figure}[h]
\centering
\includegraphics[width=.8\textwidth]{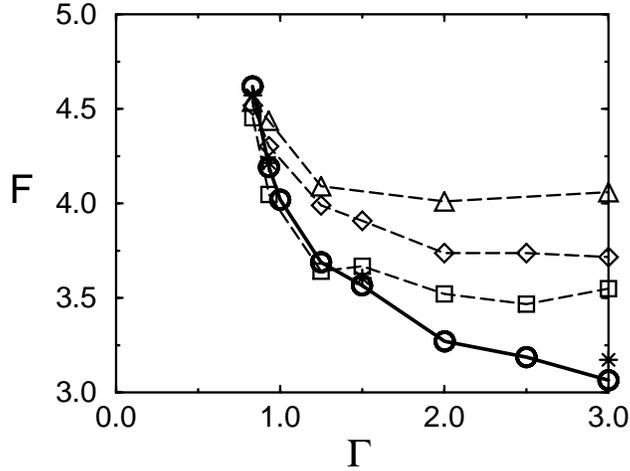}
\caption{Plot of the flatness as a function of granular temperature with
and without a lid in the system.  At low $\Gamma$ the system is nearly 2D 
and the lid has no effect. Data for $\rho = 0.532$: ($\bigcirc$) d = 1.2 mm and ($\ast$) d = 1.6
mm without lid; ($\Diamond$) d = 1.6 mm with lid.  Data for $\rho = 0.478$ 
($\triangle$) and $\rho = 0.611$ ($\Box$) with a lid and d = 1.6 mm.
}
\label{flatness} 
\end{figure}

In order to more clearly display the evolution of the distributions,
we use a simple quantitative measure of the normalized width of the
distribution, the flatness 
(or kurtosis):  

\begin{equation}
F = \frac{<v^4>}{<v^2>^2}.
\label{eq:flatness}
\end{equation}
\noindent
For a Gaussian distribution, the flatness 
is 3 and for the broader exponential distribution, the flatness is 6.  
In the absence of a lid, the flatness demonstrates
a smooth transition from non-Gaussian to Gaussian behavior as the 
acceleration is increased (Fig. \ref{flatness}), whether the smaller
(circles) or larger (stars) particles are used.  
With the lid on, the velocity distributions remain more non-Gaussian
than in the free system for identical accelerations (or identical 
granular temperatures) and 
density (diamonds).

The crossover from Gaussian to non-Gaussian behavior 
observed without the lid is therefore not simply an
effect of increasing the vertical kinetic energy of the particles, but 
rather related to the transfer of energy from the vertical to horizontal
motion in the system via collisions, the change in the density, or the
change in dimensionality of the gas.

To determine the relative contribution of density changes to the non-Gaussian
velocity distributions in the gas, the number of particles on the plate was
increased by $15\%$ and decreased by $10\%$ from the value of 
$\rho = 0.532$ and the lid was kept on.  For all accelerations, the 
flatness decreased with increased density.  This surprising result may be
related to the fact that strongly non-Gaussian 
distributions observed at low $\Gamma$ are accompanied by strong clustering 
\cite{olafsenprl}.  If the average density is increased, the larger excluded volume
means that less phase space remains for fluctuations to persist.  The fact
that increasing the density with the lid on makes the velocity distribution
more Gaussian suggests that the crossover to Gaussian observed without the
lid is not due to the decrease in density of the gas.  

\begin{figure}[h]
\centering
\includegraphics[width=.8\textwidth]{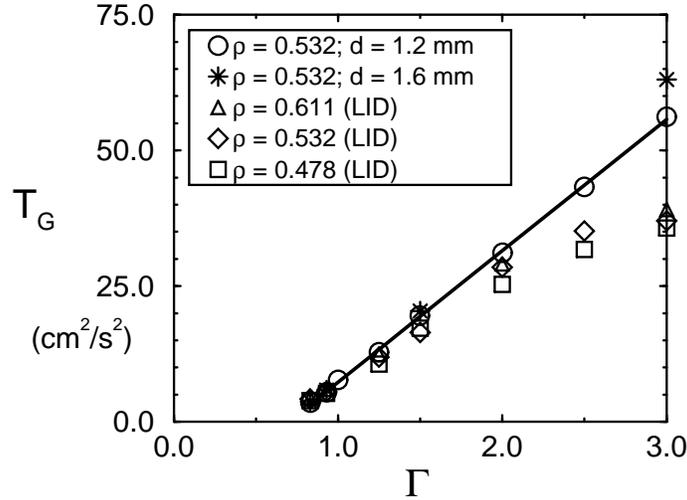}
\caption{Plot of the horizontal granular temperature, T$_G$, as a function 
of dimensionless acceleration, $\Gamma$, with and without a lid.
Without the lid ($\bigcirc$,$\ast$) the system becomes three
dimensional at high accelerations.  
}
\label{tvsg} 
\end{figure}

In addition to providing velocity distribution functions, measurements
of the particle velocities can be used to calculate the granular
temperature, T$_G=(1/2)<v^2>$.  Figure \ref{tvsg} shows the granular
temperature versus plate acceleration for several different
experimental configurations. At low acceleration ($\Gamma \leq 1$),
very few particles strike the lid and the lid has no
significant effect.  At larger $\Gamma$ it is clear that the
horizontal granular temperature is reduced when a significant number of
particles strike the lid. It is interesting to
note that T$_G$ approaches zero linearly at finite $\Gamma$,
indicating that the relationship between the driving and the
horizontal granular temperature is of the form T$_G \propto \Gamma -
\Gamma_c$. Existing models for the scaling of T$_G$ with $\Gamma$
predict a relationship of the form T$_G \propto \Gamma ^ \theta$
\cite{McNamaraL98,huntley,falcon}.  Those results are only valid at relatively
large accelerations, and clearly do not correctly describe the
behavior shown in Fig. \ref{tvsg}.

\subsection{Density-Velocity cross-correlations}
\label{crosscor}

Some insight into the origins of the non-Gaussian velocity distributions
can be obtained by investigating the relationship between the local
{\em fluctuations} in density and kinetic energy.
Puglisi {\em et al.} \cite{puglisi} have proposed a model which relates strong 
clustering to non-Gaussian velocity distributions in a driven granular
medium.  In their framework, at each local density the velocity distributions 
are Gaussian, and the non-Gaussian behavior arises from the relative weighting
of the temperature by local density in the following manner:

\begin{equation}
P(v) = \sum_{N} n(N) e^{-(\frac{v^2}{v_0^2(N)})}
\label{eq:density}
\end{equation}
\noindent
where $N$ is the number of particles in a box,  
$v_0^2(N)$ is the second moment of the distribution of velocities for
boxes with $N$ particles, and   
$n(N)$ is the number of boxes that contain N particles.  In this model, the local temperature
is a decreasing function of the local density, and the velocity distributions
conditioned on the local density are Gaussian. 

\begin{figure}[h]
\centering
\includegraphics[width=.7\textwidth]{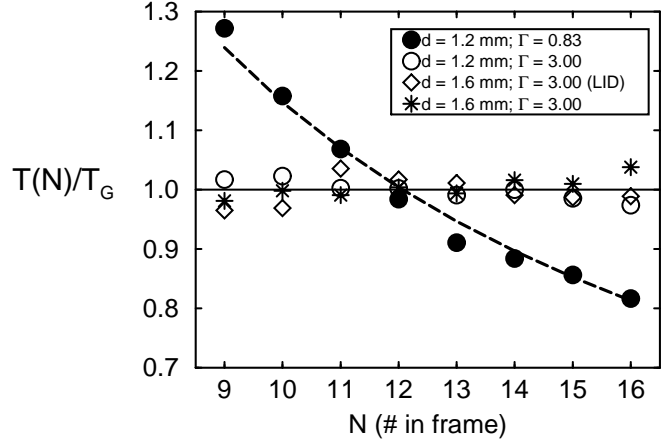}
\caption{Plot of $T(N)/T_G$ where N is the number of particles in the
camera frame (see text).  At low $\Gamma$, the local temperature and density
are strongly correlated.
At high $\Gamma$, the local granular temperature is independent
of density, even when the velocity distributions remain non-Gaussian in the
constrained system. $\rho = 0.532$. The area of the camera frame was approximately 24
ball areas.
}
\label{fig:crosscor}
\end{figure}

We have investigated the velocity distributions conditioned on 
the local density by
examining the distribution of velocities for data at a constant
number of particles in the frame of the camera.  
In the strongly clustering regime, we do observe a direct correlation
between local density and temperature.  Figure \ref{fig:crosscor} 
is a plot of the local temperature as a function of particle number in the
camera frame normalized by the global granular temperature.

The result at low accelerations is similar to the model of Puglisi
{\em et al.}  \cite{puglisi}: When the particle-particle
correlations are strongest (and larger than those of an equilibrium
hard sphere gas \cite{olafsenprl}), there is a density dependence to
the granular temperature (filled circles).  At $\Gamma = 3$, where all
of the particles are essentially uncorrelated in a 3D volume in the
absence of a lid, there is no density dependence (open circles,
stars).  However, even in the confined system at $\Gamma = 3$, where
the distribution is still not Gaussian, no appreciable density
dependence is observed (diamonds), suggesting that the non-Gaussian
velocity distributions and density-dependent temperature are not as
simply related as they are in the model of Puglisi
{\em et al.} \cite{puglisi}.  In fact, while there is a clear density
dependence on the local temperature at low $\Gamma$, the measured
velocity distribution conditioned on the local density is not
Gaussian.  At each density, the conditioned velocity distribution function is
almost identical to that of the whole: when the entire distribution is
non-Gaussian, the distribution at a single density is non-Gaussian,
and when the whole is Gaussian, each conditional velocity distribution
is Gaussian.

A more general form of Eq. \ref{eq:density} represents the total velocity 
distribution as a product of local Gaussian velocity distributions with
a distribution of local temperatures: 

\begin{equation}
P(v) = \int f(T(\vec{r},t)) e^{-(\frac{v^2}{2T(\vec{r}(t))})} d\vec{r}dt
\label{eq:tempdens}
\end{equation}
\noindent
where $T(\vec{r},t)$ is the local temperature that is varying in space and
time.  Conditioning on the local temperature would then recover
the Maxwell statistics
underlying the fluctuations.  

\begin{figure}[h]
\centering
\includegraphics[width=.8\textwidth]{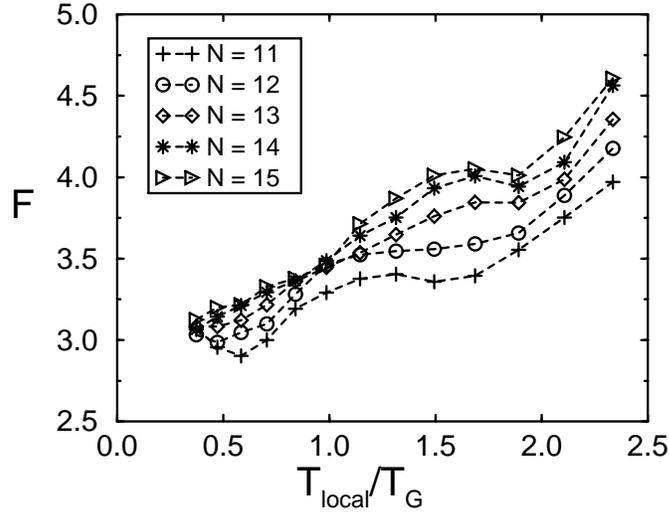}
\caption{Flatness of the velocity distribution conditioned on local
granular temperature and local density. The velocity data is separated
according to the number of particles in the frame of
the camera (N), and the local granular temperature ($T_{local}$)
determined from all of the particles in the frame.  $T_{G}$ is
the granular temperature averaged over all of the frames.  Data is for $\rho =
0.532$, $\Gamma = 0.83$, $\nu = 80$ Hz.}
\label{conditionedF}
\end{figure}

Performing this analysis on our data does not succeed in producing
Maxwell statistics.  Within small windows of local temperature, the
distributions remain non-Gaussian.  Indeed, the analysis can be
extended to condition on both the local temperature and density in the
system, but with similarly limited success except for the lowest
local temperatures, although all of
the conditioned distributions are closer to Gaussian than the full
distribution.  A plot of the flatness of the conditioned distributions
measured at $\Gamma = 0.83$ is shown in Fig. \ref{conditionedF}.  For
each density (number of particles in the frame of the camera) the
flatness is close to 3 for the 'coolest' fluctuations, but
systematically increases for larger local temperatures.  The flatness of the
full distribution for the data shown in the figure is 4.7.

If all of the measured velocities are normalized by the local granular
temperature and combined into a single distribution,
that distribution is 
Gaussian, as observed in a simulation of vibrated granular media
\cite{bizonphd}.  This is a rather surprising result, and the 
origin of this behavior is not understood.

\section{Discussion}
\label{discussion}
\subsection{Clustering and Collapse}
An understanding of the interesting dynamics displayed in both the gas
phase and the two-phase coexistence regions will require a better
understanding of the flow of energy from the plate to the granular
layer.  Energy flow into thicker layers from a vibrating surface have
been extensively studied, but the dynamics of the monolayer system are
quite different \cite{McNamaraL98,huntley,falcon}.  Energy is dissipated much more slowly than in
thicker layers, and therefore the chaotic dynamics of particles on an
oscillating plate must be considered \cite{LosertCG98,ecc}. 
The net energy transferred to the vertical motion by
the plate must balance the energy dissipated by inter-particle
collisions.  It is this balance that determines the steady-state
horizontal granular temperature. In order to generate an equation of
state, an expression for the energy input by the plate is required.
The motion of a single ball on a plate displays the characteristics of
low-dimensional chaos, but when coupled with a large number of similar
systems through inter-particle collisions, the result is apparently a
very regular rate of energy flow, producing a system that looks in
many ways very much like an equilibrium system of a large number of
interacting particles \cite{ecc}.

The interaction of the spheres with the plate, in particular the
apparent lack of any periodic or chaotic attractors with average energies
less than the period-one orbit, may partially explain the two-phase
coexistence and hysteresis observed in this system
\cite{LosertCG98}. If the kinetic energy of the spheres drops too low,
they will fall into the 'ground state', where they remain at rest on
the plate, and the energy input drops to zero.  Collisions from
neighboring spheres may keep a sphere from falling into the ground
state, but also dissipate energy.  Quantitative predictions of the
conditions under which density fluctuations can nucleate a collapse,
as well as the stability of the collapse-gas interface, may be
derivable from considerations of the sphere-plate dynamics, coupled
with the kinetic theory of dense, inelastic gases.

The appearance of strong clustering and associated nearly exponential velocity
distributions at low accelerations may also be strongly influenced by
ball-plate dynamics.  The clusters are low energy regions of the
system, and at low plate accelerations the rate of energy input may be
a strong function of average particle energies, thereby enhancing the
clustering tendency of inelastic particles.  The details of this
process are specific to our system, but all excited granular
media require external forcing, and the rate of energy input is
typically not independent of the dynamics of the grains themselves.
\subsection{Two Dimensional Granular Gas}
\label{ggdiscussion}

The granular layer at relatively high accelerations when constrained
with a lid appears to be well suited for comparison with recent
theoretical work on forced granular gases \cite{vannoije,bizon}.
The unusual behavior of the system at low accelerations appears to be
intimately tied to the interaction of the balls with the plate, and
the unconstrained system at high accelerations has a somewhat
ill-defined density.  In contrast, once the average kinetic energy of
the particles in the constrained system is large enough that the
particles can explore the entire volume of the cell,
the effective density does not vary with acceleration. 
Our results show that the correlations and velocity
distributions that are independent of $\Gamma$, and so do not appear to be
very sensitive to the details of the energy input.  (Although the way
energy is transferred from the vertical direction to the horizontal,
through collisions between particles at heights that are not too
different, may have important implications.)

As described in section \ref{velocity}, the tails of the velocity
distribution in this regime are reasonably well described by $P(v)
\propto$ exp$(-|v|^{3/2})$, in agreement with recent experimental and
theoretical results.  However, 
the details  of how energy is transferred to
the horizontal motion in the shaking experiments are sufficiently
complicated that the agreement with the theoretical calculation
of van Noije, et al.,\cite{vannoije} is perhaps surprising.  Furthermore, 
the experimental system of Losert, et al.,\cite{LosertCDKG99} was
not constrained to be two-dimensional, 
but they do not obtain the crossover to Gaussian
measured in our system without the lid.  

In the calculations of van Noije, et al., the inelastic particles are
forced by uncorrelated white noise.  In the experiment, the energy is
input into the horizontal velocities from the vertical velocity via
collisions.  Given the complex dynamics of the system, the collisional
forcing may
be reasonably well described by uncorrelated accelerations, but
the forcing occurs only during inter-particle collisions.  For
accelerations that occur much less frequently than the inter-particle
collision times, the velocity distribution should approach the freely
cooling result ($P(v) \propto$ exp$(-|v|)$) \cite{esipov}, whereas white
noise forcing is well modeled by accelerations that occur on
timescales much faster than the inter-particle collision time \cite{puglisi}. However, a recent
numerical simulation of white noise forcing \cite{bizon} found
empirically that the behavior of the system was independent of the
ratio between the rate of accelerations and the rate of collisions
as long as the ratio was greater than one.  This result suggests that
the van Noije, et al., calculations are in fact valid for the granular
monolayer, and the agreement between theory and experiment is not
simply fortuitous.

Both our experiments and the results of Losert, et al., are consistent
with velocity distributions that behave like $P(v)
\propto$ exp$(-|v|^{3/2})$ in the tails, but we obtain that result only
when the system is tightly constrained so that particles cannot pass
over one another, whereas the lid of the cell used by Losert et al.,
is 5 ball diameters above the plate.  Furthermore, Losert et al.,
include measurements up to $\Gamma = 8$,
whereas the our results for an unconstrained layer show a 
velocity distribution that becomes almost completely Gaussian when
$\Gamma$ is increased to 3.  There are several differences between the
systems (stainless steel spheres on an aluminum plate vs. glass beads
on Delrin in Losert, et al.), that might effect the results, but the
most significant difference may be that suggested by Figure \ref{overlay},
that the dynamics of the layer for a particular coverage 
are governed by $\Gamma$ and the ratio
of the driving frequency to $\nu _c =
(g/d)^{1/2}$, where $d$ is the sphere diameter. Since even our largest
particles are 2.5 times smaller than the glass beads used by Losert,
et al., our measurements at $\nu = 70$Hz are in a  different
regime than their measurements at $\nu = 100$Hz.  In particular, our
system at $\Gamma = 3$ may be more fully three
dimensional than that of Losert, et al., and $\Gamma = 8$.  Consistent
with this picture, Losert,
et al. do find that the exp$(-|v|^{3/2})$ scaling fails for lower
frequencies.  However, their measured velocity distribution does not
appear closer to Gaussian, as might be expected from our results. 

In summary, the vibrated granular monolayer has proved to be a rich
testbed for the non-equilibrium dynamics of excited granular media.
The ability to precisely measure statistical distribution functions
under well controlled experimental conditions should allow for careful
tests of the predictions of kinetic theory.  A complete
understanding of the dynamics will require a thorough study of the
phase space, as well as a more careful consideration of the dynamics of energy
transfers in the system. 

\section{Acknowledgments}
This work was supported by an award from the Research Corporation, a grant 
from the Petroleum Research Fund and grant DMR-9875529 from the
NSF.  One of us (JSU) was supported by a fellowship from the Sloan Foundation.

\end{document}